\documentclass[runningheads]{llncs}
\usepackage[T1]{fontenc}
\usepackage{multirow}
\usepackage{booktabs}
\usepackage{colortbl}
\usepackage[table,xcdraw]{xcolor}
\usepackage{hyperref}
\usepackage{graphicx}
\usepackage{geometry}
\usepackage{amsmath}
\usepackage{wrapfig}
\usepackage{makecell}
\usepackage{subcaption}
\usepackage{amssymb}

\begin{document}
\title{FIGROTD: A Friendly-to-Handle Dataset \\for Image Guided Retrieval with Optional Text}
%
%
\author{Hoang-Bao Le, Allie Tran, Binh T. Nguyen, Liting Zhou, Cathal Gurrin}
\authorrunning{Le et al.}

\institute{ADAPT Centre, School of Computing, Dublin City University, Dublin, Ireland \and
Ho Chi Minh University of Science, Vietnam National University, Ho Chi Minh City, Vietnam\\
\email{bao.le2@mail.dcu.ie, \{allie.tran, liting.zhou, cathal.gurrin\}@dcu.ie}}

\maketitle              

\begin{abstract}
Image-Guided Retrieval with Optional Text (IGROT) unifies visual retrieval (without text) and composed retrieval (with text). Despite its relevance in applications like Google Image and Bing, progress has been limited by the lack of an accessible benchmark and methods that balance performance across subtasks. Large-scale datasets such as MagicLens are comprehensive but computationally prohibitive, while existing models often favor either visual or compositional queries. We introduce \textbf{FIGROTD}, a lightweight yet high-quality IGROT dataset with 16,474 training triplets and 1,262 test triplets across CIR, SBIR, and CSTBIR. To reduce redundancy, we propose the \textbf{Variance Guided Feature Mask (VaGFeM)}, which selectively enhances discriminative dimensions based on variance statistics. We further adopt a dual-loss design (InfoNCE + Triplet) to improve compositional reasoning. Trained on FIGROTD, VaGFeM achieves competitive results on nine benchmarks, reaching 34.8 mAP@10 on CIRCO and 75.7 mAP@200 on Sketchy, outperforming stronger baselines despite fewer triplets. Our dataset and code are available at \href{https://github.com/baohl00/Friendly_IGROT_Dataset}{here}.
\end{abstract}

\section{Introduction}

Given a reference image with or without a textual description, \textbf{Image-Guided Retrieval with Optional Text (IGROT)} aims to retrieve one or more images from a collection that are visually similar to the input image and semantically aligned with the caption if it exists. This paradigm is widely used in real-world search engines such as Google Image\footnote{\href{https://images.google.com/}{https://images.google.com/}}, Google Lens\footnote{\href{https://lens.google/}{https://lens.google/}}, and Bing\footnote{\href{https://www.bing.com/}{https://www.bing.com/}}. By allowing queries to be expressed visually or visually + textually, IGROT provides users with more flexible search options, especially in scenarios where they lack precise textual descriptions. Moreover, beyond images, IGROT can serve as a stepping stone toward retrieval in more complex domains such as video (\cite{hummel2024egocvr}, \cite{rossetto2025castle}), which require spatio-temporal reasoning and human-centric understanding.  

In general, IGROT can be divided into two main settings: \textbf{without text} and \textbf{with text}.  
\textit{Without text}, IGROT reduces to classical \textbf{Visual Retrieval}~\cite{datta2008imageretrieval}, where the system must return images visually consistent with the input. With the emergence of Vision-Language Models (VLMs) such as CLIP~\cite{radford2021learning} and BLIP~\cite{li2022blip}, this task has become more robust, as these models provide joint embeddings that capture image semantics and their alignment with language.  
\textit{With text}, IGROT covers cases where users wish to apply modifications to a reference image via natural language. Two representative tasks are \textbf{Composed Image Retrieval (CIR)}~\cite{Vo_2019_CVPR}, where the reference is an image plus a relative text modification, and \textbf{Composite Sketch+Text Based Image Retrieval (CSTBIR)}~\cite{cstbir2024cstbir}, where a sketch is refined with text to query a photo. These extensions have become increasingly important as VLMs reshape how humans interact with visual search systems.  

Despite progress, two major challenges hinder further research in IGROT.  
First, the lack of an accessible, unified benchmark. Existing datasets are separated by subtask, making cross-task study difficult. Large-scale datasets such as MagicLens~\cite{zhang2024magiclens} contain tens of millions of triplets, offering diversity but at the cost of accessibility, as their size demands enormous computational and annotation resources. This limits participation, especially from students and researchers with modest resources.  
Second, through our experiments, we observe the necessity of a \textbf{shared embedding space} between query and target features. Without it, retrieval consistency degrades. To address this, we introduce a dual-loss design combining InfoNCE and Triplet loss. The latter provides complementary supervision by discouraging trivial matches to the reference image alone, thus improving compositional reasoning.  

To tackle these issues, we propose \textbf{FIGROTD}, a \textit{Friendly IGROT Dataset} containing 16,474 training triplets and 1,262 test triplets. FIGROTD is designed to be lightweight yet high-quality, making it practical for both resource-limited researchers and students. Alongside the dataset, we present a simple but effective baseline, the \textbf{Variance Guided Feature Mask (VaGFeM)}, which introduces a variance-based masking strategy to emphasise discriminative dimensions in fused representations. Our method achieves competitive performance against state-of-the-art approaches on nine diverse benchmarks, including FashionIQ~\cite{guo2019fashion}, CIRR~\cite{Liu2021cirr}, CIRCO~\cite{agnolucci2024isearle}, PatternCom \cite{psomas2024remoteSensingCIR}, Sketchy~\cite{liu2017sketchy}, TUBerlin~\cite{zhang2016tuberlin}, QuickDraw~\cite{dey2019quickdraw}, PKU \cite{pang2018pku} and the test set of FIGROTD.

In summary, the main contributions of this paper are:
\begin{itemize}
    \item We introduce \textbf{FIGROTD}, a high-quality and resource-friendly dataset for IGROT tasks, enabling fair comparison and easier experimentation across CIR, SBIR, and CSTBIR. 
    \item We propose the \textbf{Variance Guided Feature Mask (VaGFeM)}, a lightweight yet effective module that selectively emphasizes informative dimensions, yielding competitive results on six public benchmarks.
    \item We design a dual-loss training objective that combines InfoNCE with Triplet loss, demonstrating its benefits in CIR and analyzing its trade-offs in SBIR.
\end{itemize}

\section{Related Works}

\paragraph{Composed Image Retrieval (CIR), SBIR, and CSTBIR.}  
Composed Image Retrieval (CIR) has been extensively studied, where the goal is to retrieve a target image given a reference image and a textual modification. Early methods such as TIRG \cite{Vo_2019_CVPR} and ComposeAE focused on simple feature concatenation or gating mechanisms. More recent methods, including TransAgg \cite{liu2023zeroshot}, PVLF \cite{wang2025pvlf}, and CoLLM \cite{huynh2025collm}, exploit transformers or large-scale language models to better align visual and textual cues. Despite strong results, these approaches often require millions of triplets for training, making them computationally expensive and annotation-heavy.  

Sketch-Based Image Retrieval (SBIR) has a longer history, focusing on aligning sketches with natural images. Works such as ZSE-SBIR~\cite{lin2023zsesbir}, DCDL~\cite{li2025dcdl}, and CAT \cite{sain2023cat} design cross-modal encoders to handle modality gaps between sketches and real images. However, SBIR queries are purely visual and often emphasize structural cues rather than compositional semantics, making it challenging to incorporate fine-grained text modifications.  

Compositional Sketch-Text-Based Image Retrieval (CSTBIR), introduced recently in~\cite{cstbir2024cstbir}, extends this paradigm by allowing users to query with a sketch plus textual description. This setting addresses real-world cases where users cannot name an object precisely, but can visualize it through a sketch. CSTBIR unifies the strengths of CIR and SBIR, but research in this area remains limited due to the lack of accessible datasets and efficient training strategies.

\paragraph{Datasets for Retrieval.}  
The dataset landscape for multimodal retrieval is diverse, differing widely in scale and accessibility. In CIR, LaSCo~\cite{Levy2024lasco} introduced 360k compositional triplets, while TransAgg~\cite{liu2023zeroshot} provided 32k curated pairs. MTCIR~\cite{huynh2025collm} scaled this trend significantly with 3.4M synthetic triplets. MagicLens~\cite{zhang2024magiclens} further expanded the scale by generating over 36M triplets using web images and large language model captions. Although such large-scale resources are valuable, they are often impractical for researchers and students due to massive storage and computational costs.  

In SBIR, classical datasets such as Sketchy~\cite{liu2017sketchy}, TU-Berlin~\cite{zhang2016tuberlin}, and QuickDraw~\cite{dey2019quickdraw} have been widely used. While they provide strong baselines, their limited compositional supervision makes them less effective for tasks that require fine-grained reasoning.  

\paragraph{Motivation for IGROT and Our Contribution.}  
From these shortcomings, our work introduces FIGROTD under the IGROT benchmark, which provides a unified yet accessible dataset across CIR, SBIR, and CSTBIR. Unlike large-scale datasets that emphasize quantity, FIGROTD focuses on quality, efficiency, and usability, offering a friendly-to-handle resource for both researchers and students with limited computational resources. Furthermore, our proposed method, Variance-Guided Feature Mask (VaGFeM), demonstrates that competitive performance can be achieved with only 10k triplets, highlighting the importance of efficient representation learning over raw dataset scale.

\section{Dataset Construction} \label{data_construction} 

To construct FIGROTD, we integrate data from multiple sources covering CIR, SBIR, and CSTBIR subtasks in a lightweight but diverse manner. For the CIR branch, we start from a subset of the LAION dataset\footnote{\href{https://laion.ai/blog/laion-coco/}{https://laion.ai/blog/laion-coco/}}, also used in LAION\_combined \cite{liu2023zeroshot}, where we employ CLIP (ViT-L) \cite{radford2021learning} to compute similarity scores and retain only pairs above a threshold of $0.9$ \cite{agnolucci2024isearle}, ensuring reliable image–text alignment. For SBIR, we rely on the TU-Berlin dataset \cite{zhang2016tuberlin}, selecting 140 object categories with 10,979 sketches. Each sketch is randomly paired with one of several transfer prompts such as “a real image of this sketch”, “an image illustrating this sketch”, or “a real image that is similar to this sketch”, which provides diverse textual queries for sketch–image matching. For CSTBIR, we employ LLaVA-v1.6-Mistral-7B to automatically generate single-sentence captions describing each image, serving as the compositional component in training triplets. 

To construct the test set, we adopt a semi-automatic pipeline in which candidate modification texts are generated using LLaVA-NeXT-Interleave \cite{li2024llava}, trained on the INOVA Challenge 2025 dataset\footnote{\href{https://inovachallenge.github.io/ICME2025/}{https://inovachallenge.github.io/ICME2025/}} \cite{cuong2025quizzardinovachallenge2025}, and subsequently refined by \textit{annotators} who select queries that consistently satisfy both the target image group and the intended textual modification. Table~\ref{tab:dataset_stat} reports the statistics of FIGROTD: the training set contains 16,474 triplets across three subtasks with an average text length of 15.09 words, while the test set contains 1,262 triplets with shorter queries for CIR and CSTBIR (8–10 words on average). Finally, to provide a unified retrieval index, we merge the ground truths of SBIR and CSTBIR with the CIRCO index set \cite{agnolucci2024isearle}, resulting in a large-scale gallery of 126,026 images, which supports consistent evaluation across all IGROT subtasks.

\begin{table}[t] 
\centering 
\setlength{\tabcolsep}{5pt} 
\begin{tabular}{c c c @{\hskip 8pt} c c} 
\toprule 
\multirow{2}{*}{\textbf{Class}} & \multicolumn{2}{c}{\textbf{Train}} & \multicolumn{2}{c}{\textbf{Test}} \\ 
\cmidrule(lr){2-3} \cmidrule(lr){4-5} 
& \# Triplets & Text Length & \# Triplets & Text Length \\ 
\midrule 
CIR & 5,495 & 18.57 & 250 & 8.32 \\ 
SBIR & 5,099 & 17.66 & 770 & -- \\ 
SBTIR & 5,880 & 8.36 & 242 & 9.87 \\ 
\midrule 
\textit{TOTAL} & 16,474 & 15.09 & 1,262 & -- \\ 
\bottomrule 
\end{tabular} 
\caption{Statistics of three benchmarks in FIGROTD. We report the average text length for each class in the train and test splits. For the SBIR test set, transfer queries are randomly sampled from the list in Section~\ref{data_construction}.} 
\label{tab:dataset_stat} 
\end{table}

\section{Methodology}

\textbf{Motivation.}  
Retrieval tasks such as CIR, SBIR, and SBTIR demand embeddings that are both discriminative and robust across modalities. However, conventional fusion methods often introduce redundant or noisy dimensions, while existing losses may bias retrieval towards reference images without fully capturing fine-grained modifications. To address these issues, we design the \textbf{Variance Guided Feature Mask (VaGFeM)} to selectively enhance informative features, and introduce a \textbf{triplet loss} that explicitly penalizes over-reliance on reference images. Together, these components strengthen compositional understanding while maintaining shared embedding consistency across tasks.

\begin{figure}[!ht]
\begin{subfigure}[h]{0.5\textwidth}
\includegraphics[width=\textwidth]{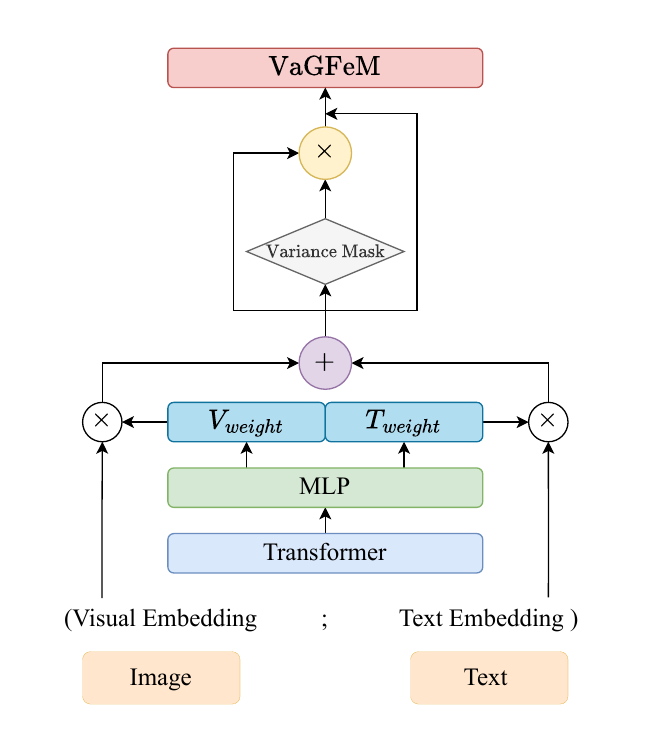}
\caption{Variance Guided Feature Mask (VaGFeM).}
\label{fig:vagfem_architecture}
\end{subfigure}
\begin{subfigure}[h]{0.5\textwidth}
\includegraphics[width=\textwidth]{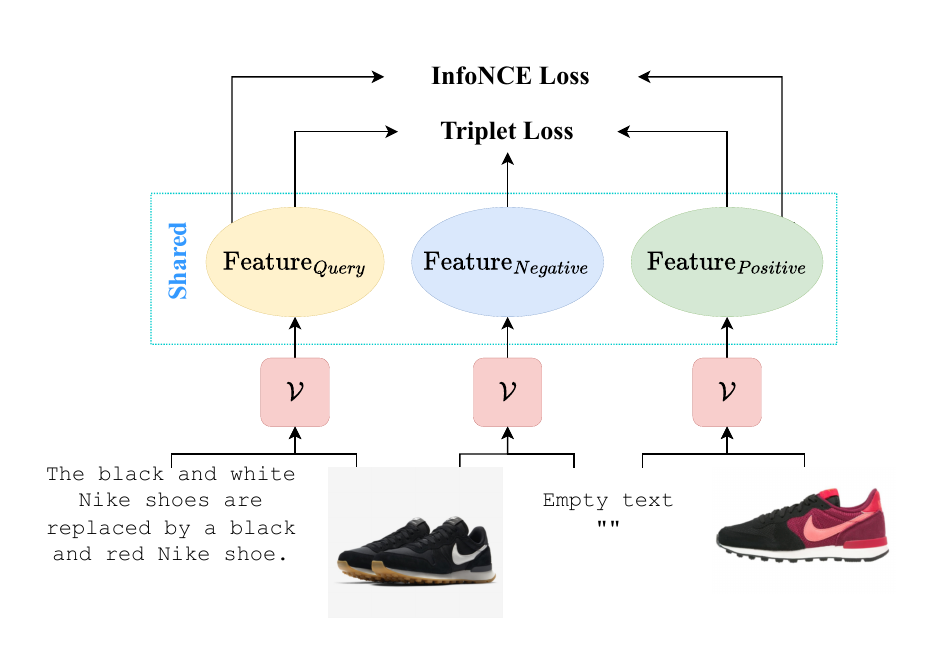}
\caption{Training with InfoNCE and Triplet Loss.}
\label{fig:vagfem_loss}
\end{subfigure}
\caption{Illustration of our proposed methodology. (a) VaGFeM enhances query features by applying variance-guided masking on fused embeddings. (b) Training employs both InfoNCE loss and triplet loss to improve compositional discrimination in a shared embedding space.}
\end{figure}

\subsection{Variance Guided Feature Mask (VaGFeM)}

To reduce redundancy and emphasise discriminative cues in joint visual-textual embeddings, we introduce the \textbf{Variance Guided Feature Mask (VaGFeM)} module in Figure \ref{fig:vagfem_architecture}. The pipeline consists of two stages: (1) constructing a union representation from image and text features, and (2) refining it using a variance-based mask.  

\paragraph{United Feature Construction.}  
Given a batch of images and corresponding texts, the image encoder produces embeddings 
$\mathbf{V} \in \mathbb{R}^{B \times D}$, and the text encoder produces embeddings 
$\mathbf{T} \in \mathbb{R}^{B \times D}$. 
Both embeddings are concatenated and passed through a lightweight transformer to capture cross-modal interactions:
\begin{equation}
\mathbf{H} = \text{Transformer}\big([\mathbf{V}; \mathbf{T}]\big) \in \mathbb{R}^{B \times 2 \times D}.
\label{eq:transformer}
\end{equation}

We split $\mathbf{H}$ into modality-specific hidden states, $\mathbf{H}_V, \mathbf{H}_T \in \mathbb{R}^{B \times D}$, 
and obtain a scalar fusion weight through a linear projection:
\begin{equation}
\boldsymbol{\omega} = \text{Linear}\big([\mathbf{H}_V; \mathbf{H}_T]\big) \in \mathbb{R}^{B \times 1}.
\label{eq:weight}
\end{equation}
The united feature is constructed as a convex combination of image and text embeddings:
\begin{equation}
\mathbf{U} = \boldsymbol{\omega} \cdot \mathbf{V} + (1 - \boldsymbol{\omega}) \cdot \mathbf{T}, 
\quad \mathbf{U} \in \mathbb{R}^{B \times D},
\label{eq:united_feature}
\end{equation}
followed by $\ell_2$ normalization to stabilise training.

\paragraph{Variance Masking.} 
To suppress uninformative dimensions, we compute the variance across the batch for each embedding dimension:
\begin{equation}
\sigma_d^2 = \operatorname{Var}(\mathbf{U}_{:,d}), \quad d = 1, \dots, D.
\label{eq:variance}
\end{equation}
We select the top-$k$ dimensions with the highest variance and define a binary mask 
$\mathbf{M} \in \{0,1\}^{B \times D}$:
\begin{equation}
\mathbf{M}_{:,d} =
\begin{cases}
1, & \text{if } d \in \text{Top-}k(\sigma^2),\\
0, & \text{otherwise.}
\end{cases}
\label{eq:variance_mask}
\end{equation}

The mask scales the embedding by its learned channel-wise activation:
\begin{align}
\boldsymbol{\alpha} &= \sigma(\mathbf{U}), \quad \boldsymbol{\alpha} \in \mathbb{R}^{B \times D}, \label{eq:sigma}\\
\mathbf{U}' &= \boldsymbol{\alpha} \odot \mathbf{M} \odot \mathbf{U} + \mathbf{U}, \label{eq:masked_feature}
\end{align}
where $\odot$ denotes element-wise multiplication.

\paragraph{Final Representation.} 
The fused representation is finally normalised:
\begin{equation}
\mathcal{V} = \frac{\mathbf{U}'}{\|\mathbf{U}'\|_2}, \quad \mathcal{V} \in \mathbb{R}^{B \times D}.
\label{eq:final_representation}
\end{equation}

This variance-guided masking ensures that only the most informative (high-variance) dimensions contribute significantly to the final query embedding $\mathcal{V}$, while redundancy is suppressed. The residual connection ($+U$) guarantees that original information is preserved, making the representation both discriminative and robust.  

\begin{figure}[!ht]
\begin{subfigure}[h]{0.5\textwidth}
\includegraphics[width=\textwidth]{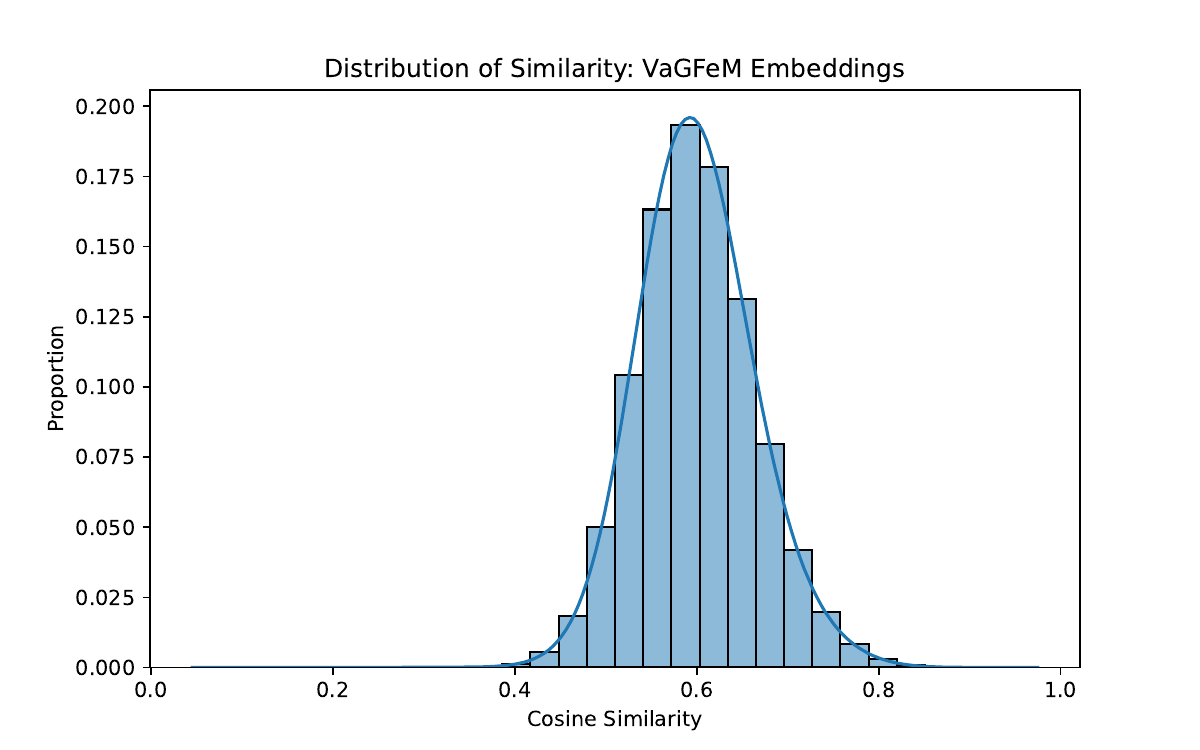}
\caption{}
\label{fig:vagfem_sim}
\end{subfigure}
\begin{subfigure}[h]{0.5\textwidth}
\includegraphics[width=\textwidth]{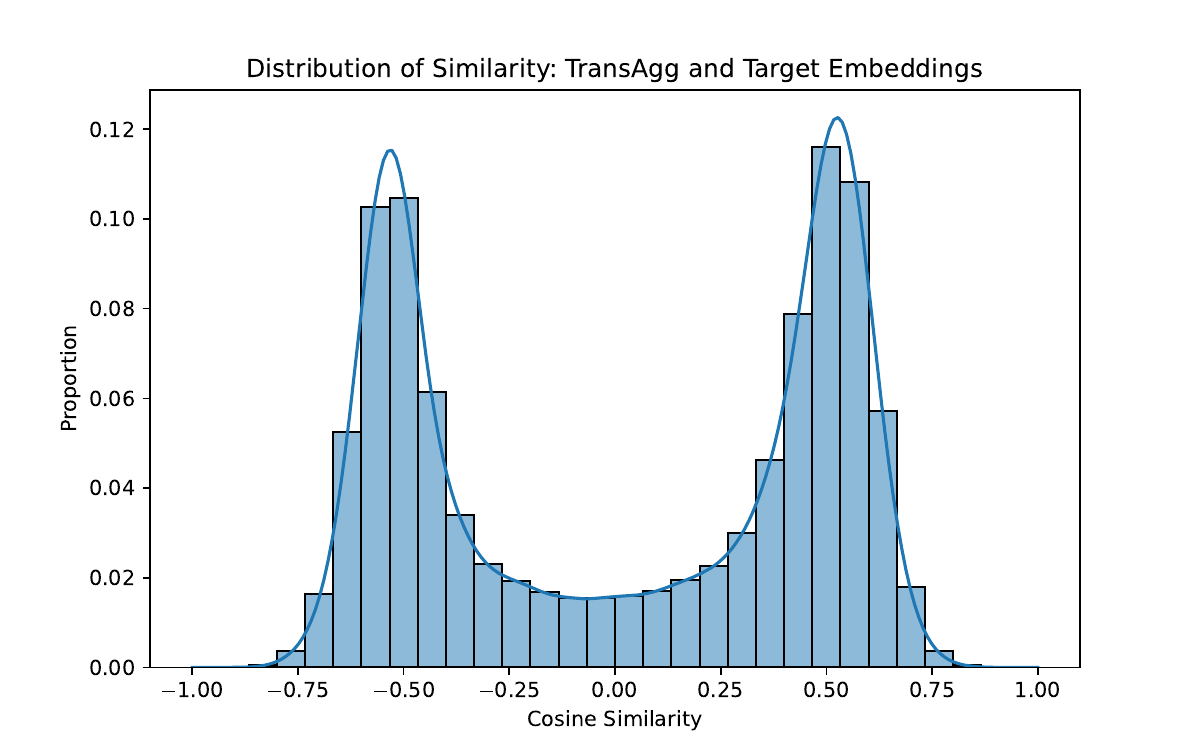}
\caption{}
\label{fig:transagg_sim}
\end{subfigure}
\caption{Comparison of cosine similarity distributions, highlighting the superior alignment of our proposed VaGFeM method (left) in Figure \ref{fig:vagfem_sim} which produces a single, high-similarity peak, against the bimodal distribution (Figure \ref{fig:transagg_sim}) of the baseline TransAgg \cite{liu2023zeroshot} model (right).}
\end{figure}

Based on the similarity distributions, VaGFeM demonstrates significantly superior alignment compared to the TransAgg \cite{liu2023zeroshot} and Target embeddings from Vision Language Model (BLIP \cite{li2022blip} in this paper). The TransAgg distribution is bimodal, indicating that while it learns to separate positive from negative pairs, it fails to bring the positive pairs into tight alignment, with their similarity peaking at a modest +0.5. In stark contrast, the VaGFeM distribution is sharply unimodal and centered at a much higher similarity score of approximately +0.6. This indicates that VaGFeM consistently maps corresponding query and target embeddings to nearly the same point in the feature space. This strong alignment is crucial for retrieval tasks, as it creates a clear and reliable margin between the correct match and all other candidates, directly translating to higher precision and top-$k$ accuracy.

\subsection{Loss Design with Triplet Loss} 

Our training framework adopts a dual-loss objective as illustrated in Figure~\ref{fig:vagfem_loss}. The first component is the \textbf{InfoNCE loss}, which encourages alignment between a query representation (reference image plus modification text) and its corresponding target image, while contrasting against all other negatives in the batch:  

$$
\mathcal{L}_\text{InfoNCE}(x,y) = \dfrac{-1}{B}\sum_{i=1}^B \log \dfrac{\exp(\mathcal{S}(x_i, y_i)/\tau)}{\sum_{j=1}^B \exp(\mathcal{S}(x_i, y_j)/\tau)} ,
$$  

where $\mathcal{S}$ denotes similarity and $\tau$ is the temperature.  

To further strengthen compositional reasoning, we introduce a \textbf{triplet loss} where the unmodified reference image (paired with an empty text prompt) is explicitly used as a hard negative. This design compels the query representation $q_i$ to be closer to its modified target $p_i$ than to the original reference $n_i$, thereby discouraging trivial matches and emphasizing semantic changes such as color, attribute, or style modifications:  

\begin{align}\label{eq:triplet_loss}
    \mathcal{L}_\text{Triplet}(x,y) = \dfrac{1}{B}\sum_{i=1}^B \max\big(0, \|q_i - p_i\|^2 - \|q_i - n_i\|^2 + \alpha \big),
\end{align}  

where $\alpha$ is the margin.  

The final training objective combines both components:  

\begin{align}\label{eq:final_loss}
    \mathcal{L}_\text{final} = \mathcal{L}_\text{InfoNCE} + \lambda \mathcal{L}_\text{Triplet}.    
\end{align}

This dual-loss formulation has distinct benefits and trade-offs. On CIR benchmarks, the triplet loss substantially boosts performance by sharpening sensitivity to subtle modifications, preventing the model from over-relying on the reference alone. On the other hand, in SBIR settings, where sketches often share structural similarity with their paired references, treating the reference as a hard negative can weaken alignment, leading to slight drops in performance. Overall, the triplet loss complements InfoNCE by enhancing compositional discrimination, though it introduces a trade-off between CIR gains and SBIR robustness (Details in Table \ref{tab:main_result} and Figure \ref{fig:triplet_loss}).

\section{Experiments}

In this section, we conducted experiments and compared the results in two different ways. Firstly, we compare the model performance with other baselines. Secondly, to present the efficient of the proposed loss function, we show itself results with and without $\mathcal{L}_\text{Triplet}(\cdot, \cdot)$. Moreover, we also show the impact of data in the ascending order of data amount. 

\subsection{Dataset} 

We select several benchmark datasets in Composed Image Retrieval: 1) \textbf{FashionIQ}: The FashionIQ dataset \cite{guo2019fashion} contains 2,005 triplets covering three fashion categories (Dress, Shirt, and Toptee) and 5,179 images in the image pool. 2) \textbf{CIRR}: The CIRR dataset \cite{Liu2021cirr} comprises 4,148 image-caption input pairs targeting 2,316 images. 3) \textbf{CIRCO}: The CIRCO dataset \cite{agnolucci2024isearle} includes 800 queries and a 123,403-image target collection. 4) \textbf{PatternCom}: The PatternCom dataset \cite{psomas2024remoteSensingCIR} has 21,571 queries across 6 attributes in Remote Sensing domain and 30,400 images in the target collection. In Sketch-Based Image Retrieval domain: 1) \textbf{Sketchy}: The Sketchy dataset \cite{yelamarthi2018zssbir} has 12,694 queries including 21 classes from ImageNet-1k and 12,694 target images. 2) \textbf{TUBerlin}: The TUBerlin dataset  \cite{zhang2016tuberlin} with 2,400 sketches across 30 categories and 27,989 index images. 3) \textbf{QuickDraw}: The QuickDraw dataset \cite{dey2019quickdraw} consists of 92,291 queries of 30 classes and a collection of 54,146 images. 4) \textbf{PKU-Sketch-ReID} \cite{pang2018pku} contains 200 sketch images and 400 identification photos (2 photos for each sketch). 


\subsection{Experiment Setting}

Our framework is implemented with Pytorch. We follow the TransAgg setups \cite{liu2023zeroshot} and use the transformer-based 2 layer fusion module with 8 heads and GELU activation. For visual and text encoders, we utilise BLIP with ViT-B \cite{li2022blip} as it gains the best performance in TransAgg's experiments. In the UNION architecture, we use a 2-layer transformer architecture similar to the T5 Transformer \cite{radford2021t5} with 2 layers, 8 attention heads for BLIP with each head having $64$ dimensions. In the loss function, we set $\tau = 0.01$ similar to TransAgg \cite{liu2023zeroshot} settings. For a fair comparison, all experiments use $224\times 224$ images to ensure, and additionally, we only choose the results used the aforementioned pre-trained models. We keep top-$20\%$ of dimension $D$ in Equation $\ref{eq:variance_mask}$. In the triplet loss \ref{eq:triplet_loss} and final loss function \ref{eq:final_loss}, we set $\alpha = 0.3$ and $\lambda = 0.2$.

The model is optimised with the AdamW \cite{loshchilov2017adamw} optimiser with a weight decay of $1e^{-2}$. All experiments are conducted with 2 epochs using learning rate $1e^{-4}$ and a batch size of $32$ on one NVIDIA A100 80GB. For the generated captions in LlavaSCo, we adapt the vision language model LLaVA-v1.6-Mistral-7B\footnote{\href{https://huggingface.co/llava-hf/llava-v1.6-mistral-7b-hf}{https://huggingface.co/llava-hf/llava-v1.6-mistral-7b-hf}}. The checkpoint of LLaVA-NeXT-Interleave\footnote{\href{https://huggingface.co/lmms-lab/llava-next-interleave-qwen-7b}{https://huggingface.co/lmms-lab/llava-next-interleave-qwen-7b}} was trained on the given data in INOVA Challenge 2025\footnote{\href{https://inovachallenge.github.io/ICME2025/}{https://inovachallenge.github.io/ICME2025/}}.

\subsection{Experimental Results and Discussion}

\subsubsection{Main result.}
We compare our methodology with TransAgg \cite{liu2023zeroshot} but trained on our dataset FIGROTD and evaluated on its testset. We report mean Average Precision (mAP) for each class (CIR, SBIR, CSTBIR). Besides, in CIR and CSTBIR task, we add Recall at $K$ (R@K) following the metrics used in FashionIQ and in SBIR task, we report Precision at $K$ (P@K) as in other datasets.

\begin{table*}[!h]
    \centering
    \begin{tabular}{|c|c|c|c|c|c|c|c|}
        \hline \multirow{2}{*}{\textbf{Method}} & \multicolumn{2}{c|}{\textbf{CIR}} & \multicolumn{2}{c|}{\textbf{SBIR}} & \multicolumn{2}{c|}{\textbf{CSTBIR}} & \multirow{2}{*}{\textbf{mAP@100}$_\text{Average}$} \\
        \cline{2-7} & \textbf{mAP@100} & \textbf{R@10} & \textbf{mAP} & \textbf{P@10} & \textbf{mAP@100} & \textbf{R@10} &   \\
        \hline 
        TransAgg$_\text{Original}$ & 3.49 & 5.85 & 0.03 & 0.00 & 0.82 & 1.45 & 1.45 \\
        \hline 
        TransAgg & \textcolor{red}{33.83} & \textcolor{red}{47.57} & 43.88 & 46.16 & 4.25 & 9.30 & \textcolor{blue}{27.32} \\
        TransAgg$_\text{Triplet}$ & 5.09 & 8.86 & \textcolor{red}{54.89} & \textcolor{red}{58.34} & \textcolor{blue}{4.82} & 7.78 & 21.60\\

        
        \hline 
        VaGFeM & \textcolor{blue}{31.52} & \textcolor{blue}{47.17} & \textcolor{blue}{51.38} & \textcolor{blue}{54.55} & \textcolor{red}{5.96} & \textcolor{red}{11.26} & \textcolor{red}{29.63}\\
        VaGFeM$_\text{Triplet}$ & 19.47 & 29.26 & 46.33 & 49.68 & 4.34 & \textcolor{blue}{9.71} & 23.38 \\
        \hline
    \end{tabular}
    \caption{Comparison of our method against existing frameworks on three tasks of FIGROTD. \textcolor{red}{Red} numbers indicate the best results, \textcolor{blue}{Blue} ones indicate the second best.} 
    \label{tab:main_result}
\end{table*}


        

The results in Table~\ref{tab:main_result} show that our Variance-Guided Feature Mask (VaGFeM) achieves the best overall performance on FIGROTD. Compared to TransAgg, which excels mainly on CIR (33.83 mAP@100, 47.57 R@10), VaGFeM delivers more balanced results with strong gains in SBIR (\textbf{51.38} mAP, \textbf{54.55} P@10) and CSTBIR (\textbf{5.96} mAP@100, \textbf{11.26} R@10), leading to the highest average score of \textbf{29.63} mAP@100. Incorporating triplet loss improves SBIR further (e.g., 54.89 / 58.34 for TransAgg$_\text{Triplet}$) but significantly reduces CIR (5.09 vs. 33.83), indicating that false negatives from reference-as-negative sampling harm compositional reasoning. Overall, VaGFeM provides the most robust trade-off across CIR, SBIR, and CSTBIR. It can be explained that in the Triplet Loss Function \ref{eq:triplet_loss}, the negative image is not the input sketch so it makes the model understand the negative sample wrongly.

\subsubsection{Comparison with Other Baselines.}

In the zero-shot CIR benchmarks, we compare VaGFeM with strong baselines including TransAgg~\cite{liu2023zeroshot}, CoLLM~\cite{huynh2025collm}, and PVLF~\cite{wang2025pvlf}, all of which use BLIP$\text{base}$ as the backbone. As shown in Table~\ref{tab:cir_overall}, VaGFeM trained with only \textbf{10K triplets} achieves competitive or superior results across benchmarks. On CIRCO, VaGFeM reaches \textbf{34.8} mAP@10 and \textbf{39.0} mAP@50, outperforming both TransAgg (32.2 / 36.2) and TransAgg$_\text{FIGROTD}$, which relies on more data (\textbf{16K triplets}) but achieves only 19.7 / 23.1. Similarly, on PatternCom, VaGFeM improves to 25.9 mAP compared to 18.7 from TransAgg$_\text{FIGROTD}$.

While CoLLM attains stronger results on FashionIQ and CIRR by leveraging millions of triplets (e.g., 34.6 / 56.0 on FashionIQ and 78.6 / 94.2 on CIRR), VaGFeM remains highly competitive with two orders of magnitude fewer annotations. This highlights the effectiveness of variance-guided feature masking for efficient learning. Finally, although FIGROTD contains 16K triplets, our experiments show that performance already peaks with 10K, underscoring the robustness and data efficiency of VaGFeM under limited supervision.

\begin{table*}[!h]
    \centering
    \resizebox{\linewidth}{!}{
    \begin{tabular}{|c|c|c|c|c|c|c|c|c|c|c|}
        \hline \multirow{2}{*}{\textbf{Method}}& \multirow{2}{*}{\textbf{\#Triplets}} & \multicolumn{2}{c|}{\textbf{FashionIQ (R)}} & \multicolumn{2}{c|}{\textbf{CIRR (R)}} & \multicolumn{2}{c|}{\textbf{CIRCO (mAP)}} & \multicolumn{2}{c|}{\textbf{PatternCom (mAP)}}\\
        \cline{3-10} & & \textbf{@10} &  \textbf{@50} & \textbf{@10} & \textbf{@50} & \textbf{@10} & \textbf{@50} & \textbf{@200} & \textbf{@all}\\
        \hline PVLF \cite{wang2025pvlf} & 32K & \textcolor{red}{35.7} & \textcolor{red}{56.9} & \textcolor{red}{82.4} & \textcolor{red}{95.4} & - & - & - & - \\
        \hline CoLLM \cite{huynh2025collm} & 3.4M & \textcolor{blue}{34.6} & \textcolor{blue}{56.0} & \textcolor{blue}{78.6} & \textcolor{blue}{94.2} & 20.4 & 23.1 & - & -\\
        \hline TransAgg \cite{liu2023zeroshot} & 32K & 34.4 & 55.1 & 77.9 & 93.4 & 32.2 & 36.2 & \textcolor{red}{37.4} & \textcolor{red}{26.6}  \\
        \hline TransAgg$_\text{FIGROTD}$ & 16K & 25.4 & 45.2 & 71.6 & 91.9 & 19.7 & 23.1 & 19.6 & 18.7 \\
        \hline VaGFeM & 10K & 27.3 & 47.7 & 76.9 & 93.3 & \textcolor{blue}{33.3} & \textcolor{blue}{37.4} & 20.9 & 19.0 \\
        \hline VaGFeM$_\text{Triplet}$ & 10K & 28.7 & 48.8 & 77.1 & 92.8 & \textcolor{red}{34.8} & \textcolor{red}{39.0} & \textcolor{blue}{34.1} & \textcolor{blue}{25.9} \\
        
        \hline
    \end{tabular}
    }
    \caption{Comparison of our method against baseline on four benchmarks of ZS-CIR task. Although FIGROTD contains 16K triplets, we notice that the results reach the peak at the 10K triplet amount. While we reproduce the results of TransAgg on CIRCO and PatternCom, the others are from the original papers. \textcolor{red}{Red} and \textcolor{blue}{Blue} numbers indicate the best and second-best results.}
    \label{tab:cir_overall}
\end{table*}



We compare VaGFeM to strong ZS-SBIR baselines—DCDL \cite{li2025dcdl}, CAT \cite{sain2023cat}, IVT \cite{zhang2024ivt}, ZSE-SBIR \cite{lin2023zsesbir}, MagicLens \cite{zhang2024magiclens}—and to TransAgg trained on $\texttt{FIGROTD}$ , all evaluated in Table \ref{tab:sbir_overall}. Despite using only 16k pairs with a BLIP$_\text{base}$ backbone, VaGFeM and its triplet variant are highly competitive against methods trained with far more data (e.g., 57k/236k for DCDL/CAT and 36.7M pairs for MagicLens). On Sketchy, VaGFeM$_\text{Triplet}$ attains the best mAP@200 = 75.7, while plain VaGFeM is second (74.0); DCDL has the top P@200 = 76.9, but requires orders of magnitude more supervision. On TU-Berlin, VaGFeM sets the new best with mAP = 64.1 and P@100 = 74.7, surpassing DCDL (63.4/74.1) and MagicLens (62.9/73.1). On QuickDraw, our models are competitive (23.2–24.8 mAP; 34.3–34.4 P@200) but trail the best mAP of DCDL (33.6) and the best P@200 of CAT (38.8), reflecting the larger domain gap of doodle-style sketches. Finally, on PKU-Sketch, our approach leads the board with rank-1 = 28.5 and strong mAP (25.2–25.9), clearly outperforming TransAgg$_\text{FIGROTD}$ (20.0/20.2). Overall, VaGFeM demonstrates robust generalisation across diverse sketch domains, and the triplet variant further boosts Sketchy and PKU-Sketch, while incurring only minor trade-offs on TU-Berlin/QuickDraw—highlighting an effective accuracy–data-efficiency balance.

\begin{table*}[!ht]
    \centering
    \resizebox{\linewidth}{!}{
    \begin{tabular}{|c|c|c|c|c|c|c|c|c|c|c|}
        \hline \multirow{2}{*}{\textbf{Method}}& \multirow{2}{*}{\textbf{Backbone}} & \multirow{2}{*}{\textbf{\# Pairs}} & \multicolumn{2}{c|}{\textbf{Sketchy}} & \multicolumn{2}{c|}{\textbf{TU-Berlin}} & \multicolumn{2}{c|}{\textbf{QuickDraw}} & \multicolumn{2}{c|}{\textbf{PKU-Sketch}} \\
        \cline{4-11} & & & \textbf{mAP@200} &  \textbf{P@200} & \textbf{mAP} & \textbf{P@100} & \textbf{mAP} & \textbf{P@200} & \textbf{rank-1} & \textbf{mAP}\\
        \hline DCDL \cite{li2025dcdl} & CLIP-B & 57K/15K/236K & 72.6 & \textcolor{red}{76.9} & \textcolor{blue}{63.4} & \textcolor{blue}{74.1} & \textcolor{red}{33.6} & 29.6 & - & - \\
        \hline CAT \cite{sain2023cat} & CLIP-B & 57K/15K/236K & 71.3 & 72.5 & 63.1 & 72.2 & 20.2 & \textcolor{red}{38.8} & - & -\\
        \hline IVT \cite{zhang2024ivt} & ViT-B & 57K/15K/236K & 61.5 & 69.4 & 55.7 & 62.9 & \textcolor{blue}{32.4} & 16.2 & - & - \\
        \hline ZSE-SBIR \cite{lin2023zsesbir} & ViT-L & 57K/15K/236K & 52.5 & 62.4 & 54.2 & 65.7 & 14.5 & 21.6 & - & -\\
        \hline MagicLens \cite{zhang2024magiclens} & CLIP-L & 36.7M & 68.2 & \textcolor{blue}{75.8} & 62.9 & 73.1 & 15.1 & 20.4 & - & -\\
        \hline TransAgg$_\text{FIGROTD}$ & BLIP-B & 16K & 73.4 & 70.9 & 63.1 & 73.4 & 24.9 & \textcolor{blue}{34.8} & 20.0 & 20.2 \\ 
        \hline VaGFeM & BLIP-B & 16K & \textcolor{blue}{74.0} & 71.9 & \textcolor{red}{64.1} & \textcolor{red}{74.7} & 24.8 & 34.4 & \textcolor{blue}{28.0} & \textcolor{red}{25.9}\\
        \hline VaGFeM$_\text{Triplet}$ & BLIP-B & 16K & \textcolor{red}{75.7} & 72.9 & 62.4 & 73.3 & 23.2 & 34.3 & \textcolor{red}{28.5} & \textcolor{blue}{25.2}\\ 
        \hline
    \end{tabular}
    }
    \caption{Comparison of our method against existing frameworks on four benchmarks of ZS-SBIR task. Except ours, the others are trained on their own training sets. \textcolor{red}{Red} and \textcolor{blue}{Blue} numbers indicate the best and second-best results.} 
    \label{tab:sbir_overall}
\end{table*}

\subsubsection{Impact of Triplet Loss.}
Figure \ref{fig:triplet_loss} compares the baseline InfoNCE objective with the combined InfoNCE + Triplet formulation. On the CIR benchmarks (FashionIQ, CIRR, CIRCO, and PatternCom), incorporating triplet loss consistently improves retrieval accuracy. For example, FashionIQ rises from 26.35 to \textbf{29.26} R@10, CIRCO increases from 32.42 to \textbf{33.97} mAP@10, and PatternCom improves substantially from 19.65 to \textbf{25.0} mAP. These gains highlight the effectiveness of triplet loss in encouraging fine-grained compositional reasoning, as the negative sample (reference image with empty text) prevents the model from overly relying on the reference alone.  

In contrast, the SBIR datasets show mixed results. While Sketchy remains stable (74.86 $\rightarrow$ 75.03), TU-Berlin and QuickDraw experience small drops (62.04 $\rightarrow$ 60.79 and 24.07 $\rightarrow$ 22.53, respectively), and PKU-Sketch decreases from 24.5 to 22.76. This reflects the fact that sketches rely heavily on coarse visual cues; penalizing the reference as a negative can weaken robustness when queries lack textual modifications.  

Finally, on FIGROTD—which integrates CIR, SBIR, and CSTBIR—the overall performance decreases from 27.61 to 21.55 mAP@100 when adding triplet loss. This suggests that while triplet loss is beneficial for CIR tasks requiring subtle compositional discrimination, it introduces a trade-off by harming generalisation in sketch-based scenarios where visual-only queries dominate. In summary, triplet loss sharpens compositional understanding in CIR but comes at the cost of reduced flexibility in SBIR, highlighting the challenge of designing a unified loss across multimodal retrieval tasks.  

\subsubsection{Impact of Training Data Size.}

\begin{figure}[!h]
\begin{subfigure}[h]{0.53\textwidth}
\includegraphics[width=\textwidth]{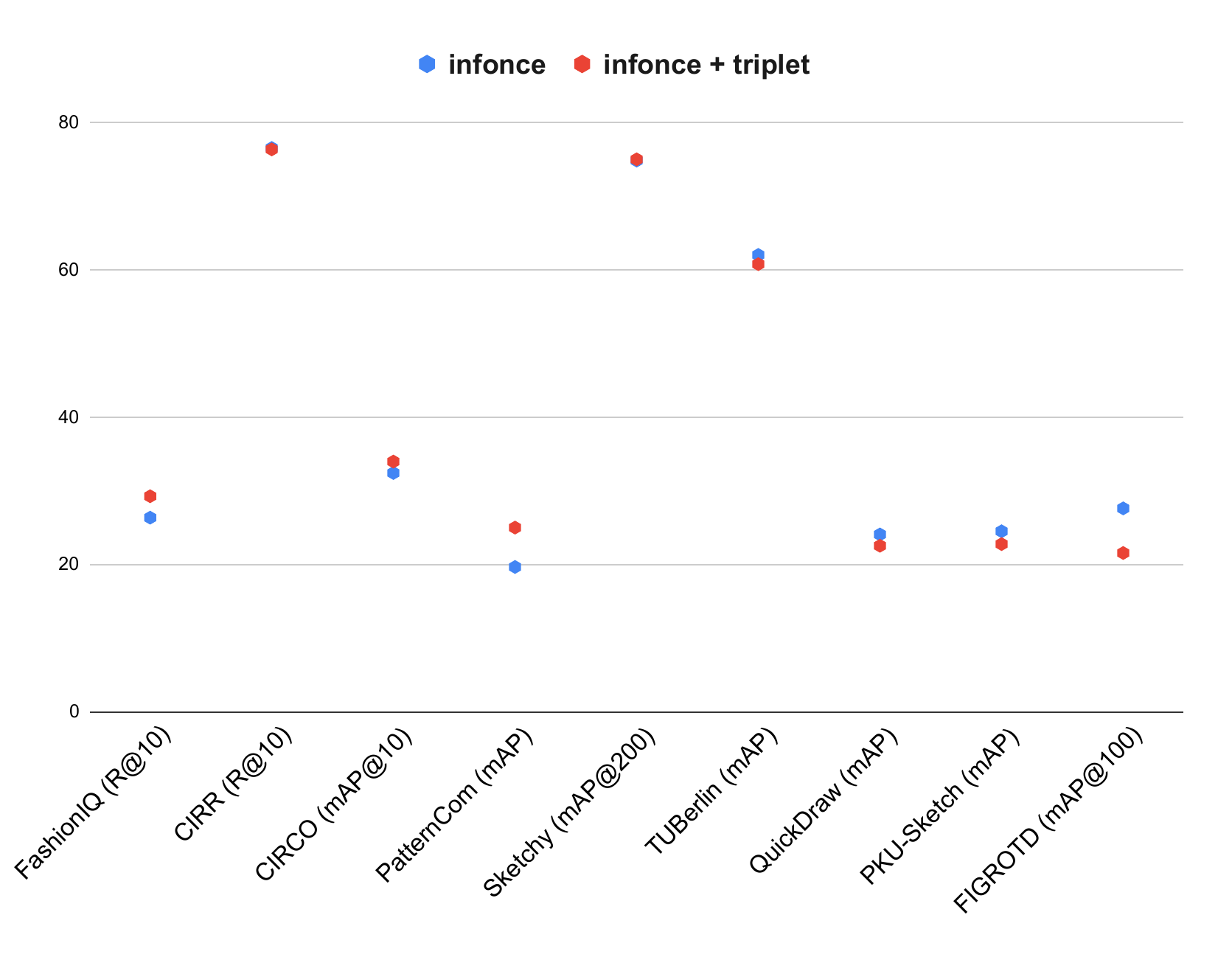}
\caption{Effect of Loss Function on Retrieval Performance}
\label{fig:triplet_loss}
\end{subfigure}
\begin{subfigure}[h]{0.46\textwidth}
\includegraphics[width=\textwidth]{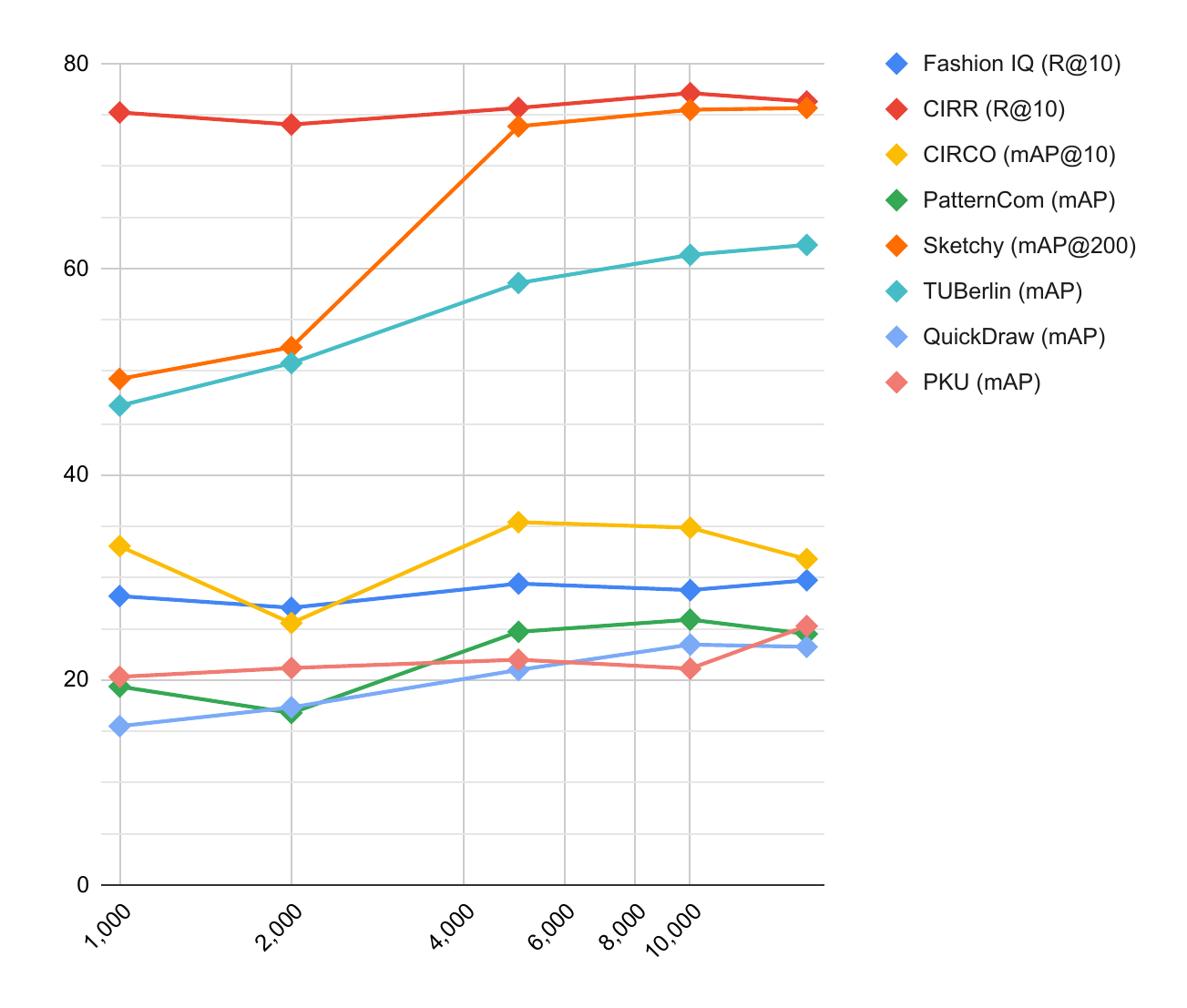}
\caption{Effect of Training Data Size on Retrieval Performance}
\label{fig:data_amount}
\end{subfigure}
\caption{Comparison of model performance across different loss functions and training data sizes on nine benchmarks. (a) Impact of loss type (infonce vs. infonce+triplet). (b) Effect of varying training data size (1K, 2K, 5K, 10K, all).}
\end{figure}

Figure~\ref{fig:data_amount} presents the performance of VaGFeM across eight benchmarks when trained with different amounts of data. Overall, the results indicate that the model reaches its peak effectiveness at around \textbf{10K triplets}, beyond which additional data offers diminishing or even negative returns. For instance, on CIRR, performance improves steadily from 75.3 at 1K to \textbf{77.1} at 10K, but drops slightly to 76.3 at 16K. Similarly, CIRCO achieves its highest score of \textbf{34.8 mAP@10} at 10K, but decreases to 31.8 at 16K. On PatternCom, performance rises from 19.3 (1K) to \textbf{25.9} (10K) before falling back to 24.5 at 16K. SBIR benchmarks such as Sketchy and TU-Berlin also plateau after 10K, with Sketchy peaking at \textbf{75.5 mAP@200} and TU-Berlin at \textbf{61.4 mAP}.  

We attribute this behavior to two factors. First, \textit{variance-guided masking} already provides strong regularisation by suppressing redundant dimensions, making the model less dependent on very large training sets. Once the key discriminative features are learned (around 10K triplets), additional noisy or weakly aligned pairs can introduce redundancy and harm generalisation. Second, tasks such as CIRCO and PatternCom are more sensitive to data imbalance, where increasing training size may emphasise easier negatives rather than truly informative contrasts, thus diluting the benefit of harder compositional reasoning.  

These results highlight that VaGFeM is not only competitive but also \textbf{data-efficient}, achieving strong generalisation with an order of magnitude fewer samples than competing methods. Training beyond 10K triplets brings limited benefit and can even slightly hurt performance, underscoring the robustness of variance-guided fusion under constrained data regimes.

\subsubsection{Qualitative Results.} 

\begin{figure}[!ht]
    \centering
    \includegraphics[width=.8\linewidth]{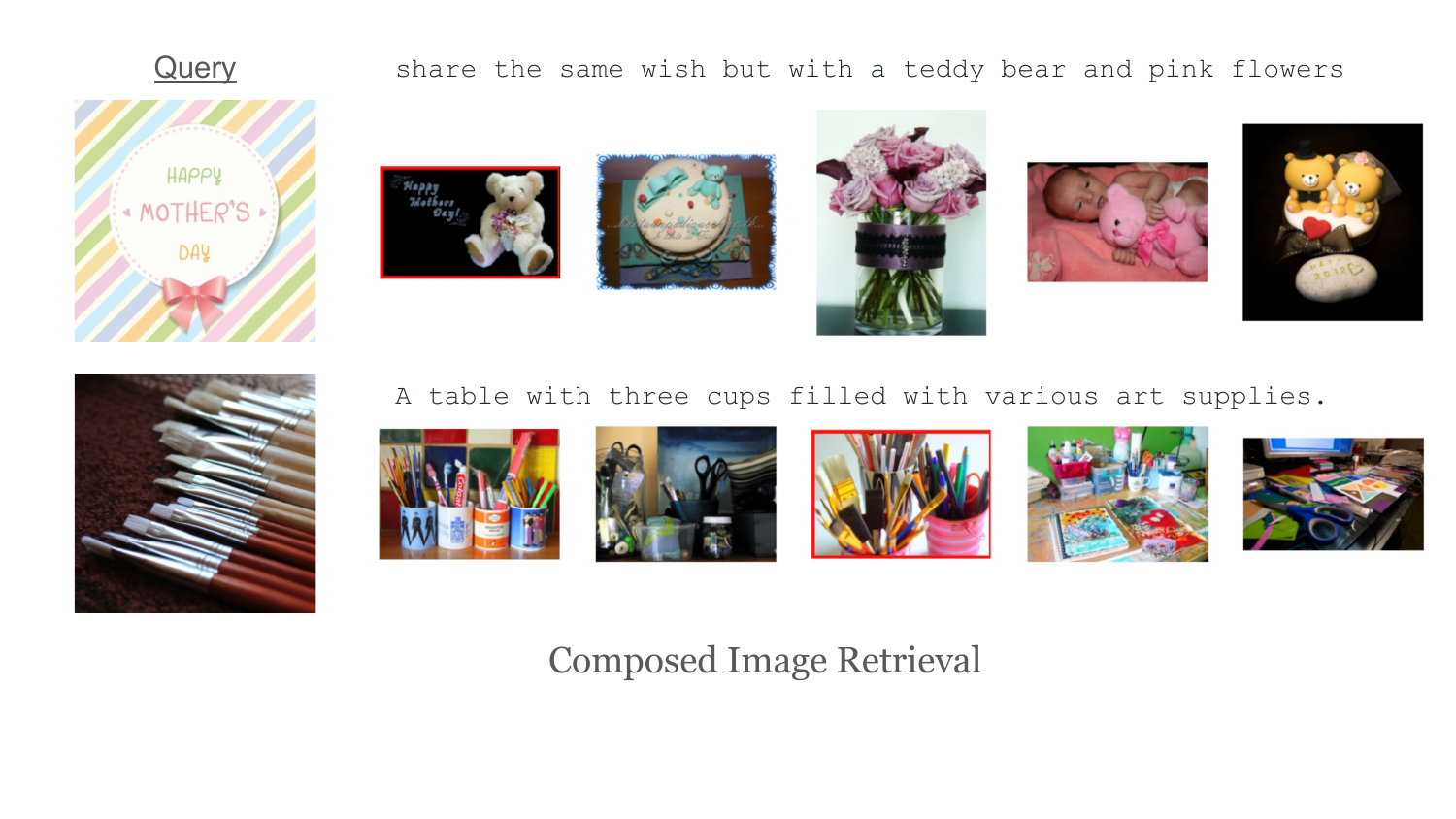}
    \includegraphics[width=.8\linewidth]{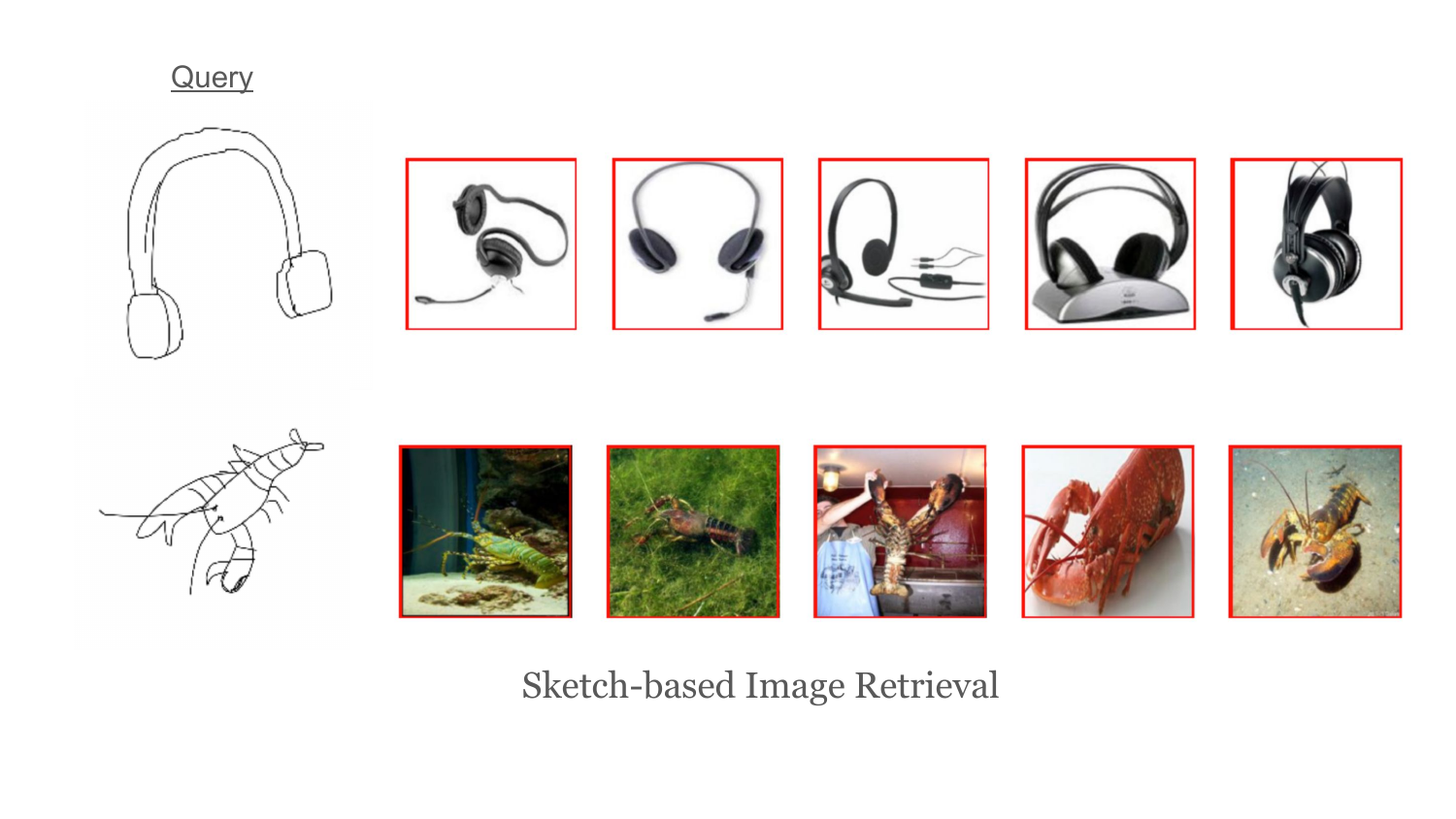}
    \includegraphics[width=.8\linewidth]{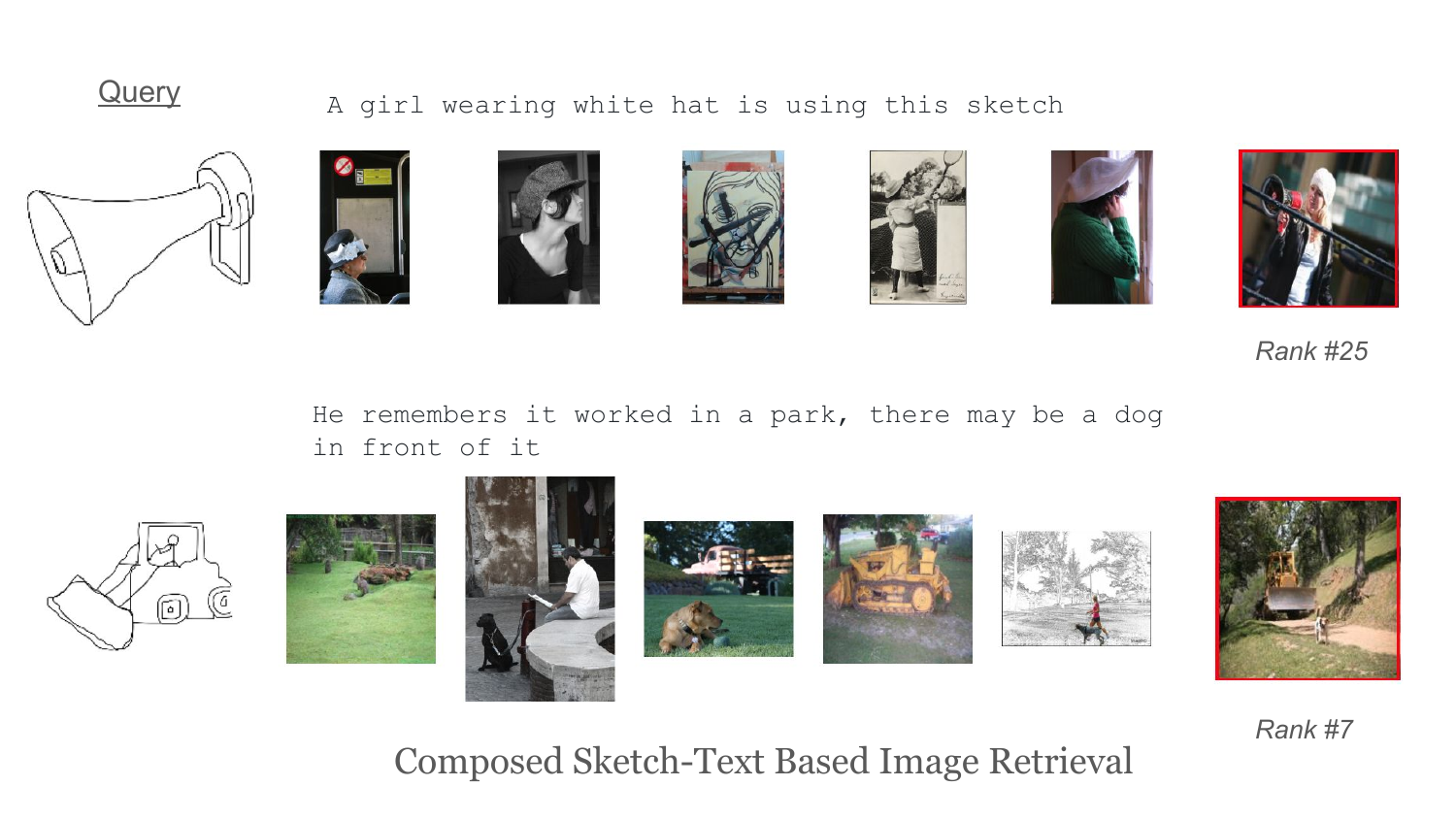}
    \caption{Qualitative retrieval examples on the three FIGROTD tasks. VaGFeM achieves strong results on CIR and SBIR, but shows limited effectiveness on CSTBIR, where fine-grained compositional reasoning is required.}
    \label{fig:qualitative_result}
\end{figure}

In Figure \ref{fig:qualitative_result}, we demonstrate qualitative retrieval examples on three tasks of FIGROTD test set (two for each class). We present the top-$5$ retrieved images, with the correct target(s) outlined in red. We hypothesise that the lack of supervised annotation can lead to the synthetic biases, and we leave the analysis of this phenomenon in the future. 

\section{Conclusion}

In this paper, we introduce FIGROTD - a friendly-to-handle Image Guided Retrieval with Optional Text that is built in a good quality for researchers to conduct experiments, apply new ideas and exploit new insights on it. We introduce VaGFeM - a simple architecture despite owning the smaller parameters, still gains competitive results. Moreover, we also perform that with additional loss function - Triplet Margin, the performance of model increases in Composed Image Retrieval Task while slightly decreases in Sketch Based Image Retrieval problem. We hope that these insights can be helpful for researchers to develop new experiments and solve the problem in the category imbalance of FIGROTD as same as improve the quality of its test set.
%
%
%
\section*{Acknowledgment}

    This publication has emanated from research supported in part by research grants from Science Foundation Ireland (SFI) under grant numbers  SFI/13/RC/2106\_P2 and 18/CRT/6223, and co-funded by the European Regional Development Fund.

\bibliographystyle{splncs04}
\bibliography{myref}

\end{document}